\documentclass{ws-procs961x669}            
\usepackage{amssymb}
\begin{document}
\newcommand{\spur}[1]{\not\! #1 \,}
\newcommand{\be}{\begin{equation}}
\newcommand{\ee}{\end{equation}}
\newcommand{\bea}{\begin{eqnarray}}
\newcommand{\eea}{\end{eqnarray}}
\newcommand{\nn}{\nonumber}
\newcommand{\dd}{\displaystyle}
\newcommand{\bra}[1]{\left\langle #1 \right|}
\newcommand{\ket}[1]{\left| #1 \right\rangle}

%
\begin{flushright}
    {BARI-TH/23-750}
\end{flushright}

\title{Novelties from Flavour Physics}

\author{F. De Fazio$^*$ }

\address{INFN, Sezione di Bari,\\
Via Orabona 4, 
Bari, I-70126, Italy\\
$^*$E-mail: fulvia.defazio@ba.infn.it}

\begin{abstract}
Our present understanding of elementary particle interactions  is the synergic result of   developments of theoretical ideas and of  experimental advances that  lead to the theory known as the Standard Model of particle physics. 
Despite the uncountable experimental confirmations, we  believe that it is not yet the ultimate theory, for a number of reasons that I will briefly recall in this lecture. My main focus will be on the role that
Flavour Physics has played in  development of the theory in its present formulation, as well as on the opportunities that this sector offers to discover physics beyond the Standard Model. 
\end{abstract}


\bodymatter

\section{Introduction}\label{intro}
The Standard Model (SM) is a gauge theory based on the group $SU(3)_c \otimes SU(2)_L \otimes U(1)_Y$ spontaneously broken to $SU(3)_c \otimes U(1)_Q$. $SU(3)_c$ is the  group of strong interactions; $SU(2)_L $,  $U(1)_Y$  the  groups of weak isospin and hypercharge, respectively,  describing electroweak interactions.  The group of electromagnetism is   $U(1)_Q$, that remains unbroken. The  theory comprises matter fields that are spin 1/2 fermions, and a single scalar, the Higgs boson. Fermions are collected in three generations, each containing quarks and leptons. The consequences of the existence of three generations are the  subject of flavour physics. 
Left-handed fermions transform as doublets under $SU(2)_L $, their right-handed counterparts as singlets. 
 Neutrinos appear only in the left-handed doublets, since they are assumed to be massless. 
In the SM   the three generations appear to be identical to each other as far as their interactions with the force carriers are concerned, since they share  the same quantum numbers under the gauge groups. This {\it universality of weak interactions} has  important  consequences. It is at the basis of the cancellation of gauge anomalies in the SM. Moreover, it leads to stringent predictions such as lepton flavour universality (LFU) recently challenged by the experiment, as will be reviewed in the following.

Before introducing the Higgs field in  the theory  and its interaction with the fermions, such an universality leads to
a global flavour symmetry  of the SM. At this stage, both fermions and gauge bosons are massless: a mass term would be incompatible with the SM  gauge symmetry. In order to preseve it, generation of the masses in the SM occurs through the Higgs mechanism. The interaction of the Higgs with the fermions is realized in the Yukawa term of the SM lagrangian density. The Yukawa matrices that appear in this term need to be diagonalized in order to identify the quark mass eigenstates. However, they cannot be all simultaneously diagonalized. A common choice is to introduce a mixing matrix  for down-type quarks, the Cabibbo Kobayashi Maskawa (CKM) matrix  that transforms weak eigenstates into mass eigenstates. Mixing of down-type quarks leads to flavour violation, occurring only in charged current interactions. Many of the observed features of weak interactions in the flavour sector are related to the properties of the CKM matrix.
This is a complex unitary matrix. Exploiting the global symmetries of the SM one finds that it has  four independent parameters,  one of which is a complex phase  responsible for  CP violation in weak decays of quarks in the SM. 
Unitarity of the CKM matrix together with the universality of $Z$ couplings to the three generations explains the absence of flavour changing neutral currents (FCNC) at tree-level.
The observed hierarchy among several quark level transitions can be understood noticing an intriguing hierarchy among the CKM elements. One of the subsequent sections will be devoted to this matrix and its elements. 

The history of the SM is full of success and experimental confirmations but it leaves behind a number of unanswered questions.
Some of these are related to cosmological observations, like the observed dominance of matter over antimatter. A  mechanism responsible for it, presumably acting  soon after the big bang, implies an amount of CP violation  that cannot be accounted for in the SM. Another one is the existence of dark matter,   motivated on astrophysical grounds, for which the SM has no viable candidate. 
The observed  hierarchy among the fermion masses has no explanation in the SM: for example more than five orders of magnitude divide the electron and the top quark masses. The SM  does not motivate the observed number of fermion generations. The  large difference between the electroweak scale, set by the Higgs vacuum expectation value, and the Planck scale, where gravity must  be taken into account, does not seem natural. The fact that gravity is not included in the SM makes its  description of the fundamental interactions incomplete.

The increase in the experimental precision has put more stringent constraints on the theory and a number of tensions between the SM  predictions and the related experimental results have emerged. In the flavour sector such tensions are usually referred to as {\it flavour anomalies} and might be a signal that flavour physics could be the right path to find evidences of physics beyond the SM (BSM).

The attempt to solve the problems of the SM and to explain flavour anomalies, is an active research  field. A new theory  extending the SM is sought for, while experiments look for signals of the existence of new particles and/or new interactions. Flavour physics offers unique opportunities in this respect.

Indeed, the search for BSM could be performed directly at colliders increasing their energy, trying to produce new particles. Since we  do not know if such new particles exist and in which range their masses lie, the question is  up to which value  the collider energy should be pushed.
However, this is not the only way to reveal the existence of new particles. Many particles have been discovered before their actual observation. This happens because they can contribute as virtual states, in particular in loop processes suppressed at tree-level. New virtual particles in the loop can modify SM predictions in a detectable way. Moreover, this could allow to be sensitive to  mass scales larger than  directly accessibile at colliders, provided that the beam intensity is large enough.
The possibility to exploit this mechanism renders  flavour physics   a promising path to the discovery of new physics (NP). 

In order to explain the observed anomalies  a suitable BSM interpretation is required. The complexity of the task is increased since most of the processes that are scrutinized involve quark decays. Quarks are bound inside hadrons, so that  nonperturbative QCD effects and the related uncertainties must be taken into account.

As a concluding remark, it is worth  recalling that flavour physics  has already paved the way to the building of the SM. 
A remarkable example is the prediction of the existence of the charm quark  suggested by the observed suppression of FCNC in rare kaon decays. The Glashow-Iliopoulos-Maiani (GIM) mechanism provided an elegant way to explain such a suppression  introducing of a fourth quark at a time when only three quarks were known. Moreover, including the virtual charm contribution to  neutral kaon oscillation amplitude, Gaillard and Lee  were able to predict the mass of the  charm quark. The prediction was   verified with the $J/\psi$ discovery in 1974. This example is  instructive, since it  shows that the existence of a particle could be inferred from its virtual contributions.

\section{Flavour Physics in the Standard Model}\label{aba:sec1}
In the previous section we have briefly  recalled the features of flavour physics in the SM, and we have stated that this sector offers the opportunity to get insights into the existence of NP and on its possible features. Now we would like to explain how this is possible. 
In particular, questions that need to be answered are:
\begin{itemize}
\item Why three generations?
\item Can we explain the observed hierarchy of masses and mixing parameters? This is the {\it flavour puzzle}: the SM does not provide such an explanation.
\item
How can we check the correctness of the CKM description of quark mixing? 
\item 
What can we still learn from FCNC processes? 
Since the suppression of these modes is a consequence of CKM unitarity together with the universality of weak interactions, this question is strictly  related to the previous one and implies looking at rare FCNC processes.
\item
Can we explain the recently observed flavour anomalies?
\end{itemize}
There are  two ways to try to answer the previous questions. One is a  {\it bottom-up} approach, mainly driven by experiment. It consists in considering the SM as an effective theory valid at the electroweak scale and investigate the features of the more general theory from which it discends, without any reference to a specific NP model. This kind of  approach is realized in the Standard Model Effective Field Theory (SMEFT) \cite{Grzadkowski:2010es}.
SMEFT assumes that NP exists at a  scale $\Lambda \gg M_W$ and that its gauge group  contains the SM one. If the SM is regarded as an effective theory at the low scale $M_W$, it is possible to write  an effective lagrangian  consisting in the SM one plus new terms written in terms of the SM fields,
 invariant under the transformations of the SM gauge group and
  suppressed by increasing powers of $1/\Lambda$:
\begin{equation}
{\cal L}_{SMEFT}={\cal L}_{SM}^{(4)}+\frac{1}{\Lambda}C_{\nu \nu}^{(5)}Q_{\nu \nu}^{(5)}+\frac{1}{\Lambda^2}\sum_k C_k^{(6)}Q_k^{(6)}+{\cal O}\left(\frac{1}{\Lambda^3}\right)\,\,\,.
\label{LSMEFT}
\end{equation}
${\cal L}_{SM}^{(4)}$ is the SM lagrangian density consisting of the kinetic terms, of the terms describing the interactions among the fermions and the gauge bosons, of the Higgs lagrangian and of the Yukawa term. Among these it is remarkable that, except for the latter term, in all the other terms accidental  symmetries are  fulfilled. Such symmetries are those that are valid only because we impose renormalizability and can be violated by the addition of non renormalizable terms.   In the SM the baryon and lepton number conservation, as well as lepton family number conservation, belong to this category. 
 Moreover,  approximate symmetries exist: in general, these are governed by a small parameter so that when it is set to zero, such symmetries become exact. 
The possibility of violating accidental and approximate SM symmetries through the new terms in Eq. \eqref{LSMEFT} is exploited to find NP.

$Q_{\nu \nu}^{(5)}$ has dimension 5, it is known  as the Weinberg operator  responsible for neutrino mass. The last sum in \eqref{LSMEFT} contains dimension 6 operators: they are relevant to describe NP in the bottom-up approach and can violate the accidental symmetries.

The second approach is  {\it top-down} in which one works out predictions in a specific BSM framework. 
Results for flavour observables and correlations among them represent a signature of the chosen model, so that comparison with experiment might  discriminate among different scenarios. Examples will be given in the following.

\section{The CKM matrix}
The elements of the CKM matrix $V_{CKM}$ are fundamental parameters of the SM,  just like the fermion masses, the gauge couplings and the parameters of the Higgs potential.  Besides being interesting {\it per se}, the  precise determination of such elements has more profound motivations. Indeed the SM description of CP violation in weak decays is related to the complex phase of this matrix, further motivating the determination of its elements. In the SM $V_{CKM}$ is unitary: checking the unitarity constraints is
a way to look for possible deviations from SM. This is an ongoing effort  since several decades, from the theory and the experimental side with precision increasing with time. This issue is so important to deserve a deeper insight.

$V_{CKM}$ can be parametrized in different ways. In the {\it standard parametrization}\cite{ParticleDataGroup:2022pth} the independent parameters are three angles $\theta_{ij}$ ($i,j=1,2,3$) with sines and cosines $s_{ij},\,c_{ij}$ chosen to be positive and the angle $\delta$ entering through the phase $e^{i\,\delta}$, with $\delta \in [0,2 \pi]$ (kaon phenomenology  constrains $\delta \in [0,\pi]$). This parametrization satisfies unitarity  exactly but it is often difficult to handle. Hence, it is  convenient to consider the Wolfenstein parametrization\cite{Wolfenstein:1983yz},  obtained expanding each element as a  series in  powers of the small parameter $\lambda=|V_{us}|$ that approximately coincides with the sine of the Cabibbo angle $\sin \theta_C \simeq 0.22$.
In this parametrization $V_{CKM}$  reads: 
\be V_{CKM}=\left( \begin{array}{ccc} V_{ud} & V_{us} & V_{ub}\\
 V_{cd} & V_{cs} & V_{cb}\\
 V_{td} & V_{ts} & V_{tb}
\end{array} \right)= 
  \left( \begin{array}{ccc}1-{\lambda^2 \over 2} & \lambda &
 A \lambda^3 (\rho-i \eta) \\ -\lambda & 1-{\lambda^2 \over 2} & A
 \lambda^2 \\
A \lambda^3 (1-\rho-i \eta) & -A
 \lambda^2 & 1 \end{array} \right) +{\cal O}(\lambda^4)
 \label{wolf} \,. \ee
The  independent parameters  are $\lambda,\,A,\,\rho,\,\eta$. The hierarchy among the elements is  apparent: $V_{ud},\,V_{cs},\,V_{tb}$ are  ${\cal O}(1)$,   $V_{us}$, $V_{cd}$ are ${\cal O}(\lambda)$, $V_{cb}$, $V_{ts}$ are ${\cal O}(\lambda^2)$ while $V_{ub}$, $V_{td}$ are  ${\cal O}(\lambda^3)$.
In this parametrization $V_{CKM}$ is  unitary  up to ${\cal O}(\lambda^4)$. When more accuracy is required, it might be necessary to include more terms in the expansion. One possibility to do this is to define: $s_{12}=\lambda$, $s_{23}=A \lambda^2$, $s_{13}e^{-i\, \delta}=A \lambda^3(\rho-i \eta)$, so this is just a change of variables between the two parametrizations.\cite{Buras:1994ec,Schmidtler:1991tv} All the elements of $V_{CKM}$ are  then expressed in terms of $\lambda,\,A,\,\rho,\,\eta$. Next one expands all the entries in powers of $\lambda$ and  defines ${\bar \rho}=\rho \left(1-\frac{\lambda^2}{2} \right)$ and ${\bar \eta}=\eta \left(1-\frac{\lambda^2}{2} \right)$. In this way the following relations hold up to order ${\cal O}(\lambda^7)$ or  higher:
$
V_{us}=\lambda; \, V_{ub}=A \lambda^3(\rho -i \eta),\,V_{cb}=A \lambda^2,\, V_{td}=A \lambda^3(1-{\bar \rho} -i {\bar \eta})$,
therefore it is  convenient to use.

Let us consider the relations stemming from unitarity, in particular the ortogonality conditions among rows/columns. In the Wolfenstein representation, these can be represented as
triangles in the complex $(\bar \rho, \bar  \eta)$ plane.
All the triangles have the same area
$A=J/2$, with $J=Im(V_{us}V_{cb}V_{ub}^*V_{cs}^*)$  called   
Jarlskog parameter.\cite{PhysRevLett.55.1039} The angles of the triangles are
 the relative phases of two adjacent sides, i.e. of a product of  CKM  elements. The task is to determine sides and angles in different ways and check the consistency of the results. In particular, verifying that the sum of the angles of the triangles is $\pi$ is a test of the unitarity of  $V_{CKM}$  and consequently of the correctness of the SM.
However, not all the triangles are suitable for this purpose.
The one  corresponding to the ortogonality constraint
\be
 V_{ub}V_{ud}^*+ V_{cb}V_{cd}^* + V_{tb} V_{td}^*=0
\label{UT}
\ee
has received more attention than all the others for several  reasons.  
Adopting the Wolfenstein parametrization  the three terms  in \eqref{UT} are all ${\cal O}(\lambda^3)$. Except for another one, all the other ortogonality conditions involve terms of different order in   $\lambda$, and in the corresponding triangles  one of the angles and one of the sides are so small that it is very difficult to measure them. Moreover, the elements involved in \eqref{UT} are those that enter in beauty hadron decays, where large CP violating effects are expected. Considering the unitarity relation \eqref{UT} and dividing it by
$|V_{cb}V_{cd}^*|$ we obtain: 
$\displaystyle{{V_{ub}V_{ud}^* \over
|V_{cb}V_{cd}^*|}+ {V_{tb}V_{td}^* \over |V_{cb}V_{cd}^*|}+1=0
}$,
so that the basis of this triangle has
unitary length. Such a triangle is  referred to as the {\it unitarity triangle} (UT), with angles often denoted $\alpha,\,\beta,\,\gamma$. The other two sides have lengths $R_u=\frac{|V_{ub}V_{ud}^*|}{|V_{cb}V_{cd}^*|}$ and $R_t=\frac{|V_{tb}V_{td}^*|}{|V_{cb}V_{cd}^*|}$. To very good approximation one has  $V_{ub}=|V_{ub}|e^{-i\gamma}$ and $V_{td}=|V_{td}|e^{-i\beta}$, while $\alpha$ is the angle opposite to the side of length 1.
 The
fact that the sides of the UT have roughly the same size  means that the phases
of the  elements entering in their expression  are the largest ones, implying
 large effects of CP violation. 
Checks of the CKM unitarity through the determination of the sides and the angles of the UT  show an optimal consistency of the various constraints. Present world averages\cite{HeavyFlavorAveragingGroup:2022wzx}  provide
$
A=0.8132\pm^{0.0119}_{0.0060}$, $ \lambda=0.22500\pm^{0.00024}_{0.00022}$, ${\bar \rho}=0.1566\pm^{0.0085}_{0.0048} $,
${\bar \eta}=0.3475\pm^{0.0118}_{0.0054} $.

Despite the very good agreement between SM predictions and experiment regarding the UT, some tensions exist.
A long standing problem is related to  the inclusive and exclusive determinations of $|V_{ub}|$ and $|V_{cb}|$, inclusive being in both cases larger than exclusive. A more recent problem is the  {\it Cabibbo anomaly} related to the unitarity of the first row. 
We discuss them in the following subsections.

\subsection{Cabibbo anomaly}
Unitarity of the first CKM row implies 
$
\Delta_{CKM}=|V_{ud}|^2+|V_{us}|^2+|V_{ub}|^2-1=0 $.
In order to check this relation one can neglect  $|V_{ub}|^2\simeq{\cal O}(10^{-5})$.
$|V_{ud}|$ is  determined from superallowed nuclear beta decays:
$|V_{ud}|=0.97367(11)_{\rm exp}(13)_{\Delta_V^R}(27)_{\rm NS}$ \cite{Bryman:2021teu};
$\Delta_V^R$ and  ${\rm NS}$ denote uncertainties due to radiative corrections.\cite{PhysRevC.102.045501}
$|V_{us}|$ is determined from $K_{\ell3}$ decays, from inclusive hadronic $\tau$ decays and from hyperon decays, with values that disagree at 3$\sigma$ level. 
Combining the various results for both $|V_{ud}|$ and $|V_{us}|$, a global fit\cite{Crivellin:2022rhw}  provides
$|V_{ud}|=0.97379 \pm 0.00025$ and $|V_{us}|=0.22405 \pm 0.00035$, corresponding to $\Delta_{\rm CKM}=-0.00151(53)$ with a deviation from unitarity at 2.8$\sigma$ level. The discrepancy among the values of $|V_{us}|$ together with the deviation of $\Delta_{\rm CKM}$ from zero is referred to as the {\it Cabibbo anomaly}.

\subsection{$|V_{ub}|$ and $|V_{cb}|$: inclusive vs exclusive determinations}
\subsubsection{Preliminary: Heavy Quark symmetries}
Heavy quarks (HQ) are defined as those with mass larger than the QCD scale parameter $\Lambda_{QCD}\simeq 200 \,\,{\rm MeV}$, roughly  the inverse of the proton radius $R\simeq 1 \,\,{\rm fm}$. 
Properties and decays of systems with a single HQ ({\it heavy-light systems}) can be sistematically treated in the   limit $m_Q \to \infty$, where $Q=c,b$ (the top is not considered since it decays before hadronizing).
In this limit  the chromomagnetic interaction
between the HQ spin ${\vec s}_Q$ and the  total angular
momentum  ${\vec s}_\ell$ of the light degrees of freedom (light quark(s) and gluons)  vanishes, being inversely proportional to $m_Q$. Therefore
these quantities  decouple, giving rise to the HQ spin symmetry. 
Moreover, since  in
the QCD lagrangian  only 
the mass term depends on the flavour,  in the large mass limit also the HQ flavour becomes irrelevant. HQ
 spin-flavour symmetry is realized in an  effective theory   called Heavy Quark Effective Theory 
 (HQET)\cite{Neubert:1993mb}. 
 The lagrangian density of HQET is derived  writing the HQ field in QCD as $Q(x)=e^{-i\,m_Q v \cdot x}[h_v(x)+H_v(x)]$ where $v$ is the HQ four velocity and $h_v=\displaystyle\frac{1+\spur v}{2}Q$, $H_v=\displaystyle\frac{1-\spur v}{2}Q$. Substituting this in the QCD  lagrangian for the HQ:  ${\bar Q}(i {\spur D} -m_Q)Q$ ($D$ is the covariant derivative in the fundamental representation) and using the equations of motion to eliminate $H_v$, one can perform an expansion in the 
 inverse HQ mass (HQE).
The leading order term defines the HQET lagrangian:
 ${\cal L}_{HQET}={\bar h}_v i \, v \cdot D h_v$.
 Other terms can be included; at ${\cal O}(1/m_Q)$ one has 
 \be
 {\cal L}={\bar h}_v i \, v \cdot D h_v+\frac{1}{2m_Q}{\bar h}_v (i \,  {\spur D}_\perp )^2 h_v+\frac{1}{2m_Q}{\bar h}_v \frac{g_s \sigma_{\alpha \beta} G^{\alpha \beta}}{2} h_v+{\cal O}\left( \frac{1}{m_Q^2} \right)
 \,\,. \label{LmQ}
 \ee
 $D_\perp^\mu=D^\mu-(v \cdot D) v^\mu$, $g_s$ is the strong coupling constant and $G^{\alpha \beta}$ the gluon field tensor.
 The first correction in \eqref{LmQ} represents the HQ kinetic energy operator,  the second one is the chromomagnetic coupling
of the HQ spin to the gluon field.
Important applications of HQET concern the spectroscopy of heavy-light hadrons as well as in their weak decays. Incorporating corrections to the asymptotic limit through the HQE allows to improve the precision of the predictions. In the following we describe  applications  to exclusive and inclusive decays of heavy-light systems.

 \subsubsection{Generalized effective Hamiltonian for $b \to {u(c)} \ell \nu$ decays}
 According to the SMEFT prescription  
 in the bottom-up approach one can write a generalized low-energy effective Hamiltonian  to describe the semileptonic transition $b \to U \ell \nu$ ($U=u,c$), extending  the SM one with D=6 four-fermion operators
 \bea
H_{\rm eff}^{b \to U \ell \nu}&=& {G_F \over \sqrt{2}} V_{Ub} \Big[(1+\epsilon_V^\ell) \left({\bar U} \gamma_\mu (1-\gamma_5) b \right)\left( {\bar \ell} \gamma^\mu (1-\gamma_5) {\nu}_\ell \right)
\nn \\
&+&  \epsilon_R^\ell \left({\bar U} \gamma_\mu (1+\gamma_5) b \right)\left( {\bar \ell} \gamma^\mu (1-\gamma_5) {\nu}_\ell \right) + \epsilon_S^\ell \, ({\bar U} b) \left( {\bar \ell} (1-\gamma_5) { \nu}_\ell \right)\label{hamilgen}   \\&+&\epsilon_P^\ell \, \left({\bar U} \gamma_5 b\right)  \left({\bar \ell} (1-\gamma_5) { \nu}_\ell \right) +\epsilon_T^\ell \, \left({\bar U} \sigma_{\mu \nu} (1-\gamma_5) b\right) \,\left( {\bar \ell} \sigma^{\mu \nu} (1-\gamma_5) { \nu}_\ell \right)
  \Big] + h.c.  \nn
\eea
 $\epsilon^\ell_{V,R,S,P,T}$ are complex  lepton-flavour dependent  couplings.  Only  left-handed neutrinos are included. $G_F$ is the Fermi constant  and  $V_{Ub}$ is the relevant CKM  element.

\subsubsection{$V_{xb}$: Exclusive determination}
Processes induced by the quark  decays $b \to U$, with $U=u,c$ allow to determine the CKM element $V_{Ub}$. In the exclusive modes the final state is fully reconstructed. For example,   $B \to \pi \ell {\bar \nu}_\ell$ and $B \to \rho \ell {\bar \nu}_\ell$ are useful  to determine $|V_{ub}|$, while $B \to D^{(*)} \ell {\bar \nu}_\ell$ can be exploited to determine $|V_{cb}|$. In the SM, the  description of exclusive modes requires a set of  form factors (FF) that define the hadronic part of the transition amplitude: $\langle H_U| {\bar U} \gamma_\mu (1-\gamma_5)b|B\rangle$. In BSM scenarios new operators $O_i$ can be present as in \eqref{hamilgen}, requiring also the  FF defining $\langle H_U| O_i|B \rangle$. Being nonperturbative quantities, FF can be determined with methods such as lattice QCD or  QCD sum rules. In some cases, there exist symmetries that provide relations among them or to fix their normalization.  Let us consider  the case of $V_{cb}$  to give an example. 

$B \to D^{(*)}$ are transitions between  heavy-light mesons, hence one can exploit the HQ symmetries. These  relate, at leading order in the HQE, the  FF that parametrize  the matrix elements $\langle H_U| {\bar U} \Gamma b|B \rangle$ to  a single function $\xi$ independently on the Dirac structure $\Gamma$. $\xi$ is   the Isgur Wise (IW) function.  In the SM $B \to D^{(*)}$ decays require 6 FF,  all replaced by $\xi$, which shows the  simplification achieved in the HQ limit. 
HQET does not allow to compute $\xi$, but  fixes its normalization at the zero recoil point: $\xi(q^2_{max})=1$.  $1/m_Q^n$ and $\alpha_s(m_Q)$ corrections can be systematically added. Remarkably, the normalization of the IW function is not affected at ${\cal O}(m_Q^{-1})$, but remains exact up to ${\cal O}(m_Q^{-2})$, a result known as the Luke's theorem.\cite{Luke:1990eg} Hence, the predictions for the spectra in the dilepton invariant mass $q^2$ close to $q^2_{max}$ are affected by a small uncertainty, so that comparison with data in such a kinematical region allows a reliable  determination of $|V_{cb}|$. 
The latest average of the results obtained with this method reads: $|V_{cb}|=(38.46 \pm 0.40_{\rm exp}\pm0.55_{\rm th}) \times 10^{-3}$.\cite{HeavyFlavorAveragingGroup:2022wzx}

\subsubsection{$V_{xb}$: Inclusive determination}
In the inclusive modes the final state is not fully specified, requiring experimental efforts. Reliable  tools  for the theory  description of these modes are available, exploiting  the optical theorem and the HQE:
we briefly explain them. 

The Hamiltonian  \eqref{hamilgen}  can be written as
\be
H_{\rm eff}^{b \to U \ell \nu}= \frac{G_F}{\sqrt{2}} V_{Ub} \sum_{i=1}^5 C_i^\ell \, J^{(i)}_M\, L^{(i)M} + h.c.\,\,\, ,\label{hnew}\ee
with  $C_1^\ell=(1 +\epsilon_V^\ell)$ and  $C_{2,3,4,5}^\ell=\epsilon_{S,P,T,R}^\ell$.   $J_M^{(i)}$ ($L^{(i)M}$) is the hadronic (leptonic) current in each operator,  $M$ are Lorentz indices contracted between $J$ and $L$. 
The   inclusive semileptonic differential  decay width of a beauty hadron  $H_b$ is written as
\be
d\Gamma= d\Sigma \, \frac{G_F^2 |V_{Ub}|^2}{4m_H} \sum_{i,j} C_i^* C_j (W^{ij})_{MN} (L^{ij})^{MN} . \label{diffwid}
\ee
  $q=p_\ell+p_\nu$  is the lepton pair momentum, and    $d\Sigma=(2\pi) d^4q \, \delta^4(q-p_\ell-p_\nu) [dp_\ell]\,[dp_\nu] $, with   $[dp]=\displaystyle\frac{d^3 p}{(2\pi)^3 2p^0}$. The leptonic tensor is  $(L^{ij})^{MN}=  L^{(i)\dagger M} L^{(j)N}$.
The hadronic tensor  is obtained using  the optical theorem: $(W^{ij})_{MN}=\frac{1}{\pi}{\rm Im}(T^{ij})_{MN}$ with
\bea
(T^{ij})_{MN}&=&i\,\int d^4x \, e^{-i\,q \cdot x} \langle H_b(p,s)|T[ J^{(i)\dagger}_M (x) \,J^{(j)}_N (0)] |H_b(p,s) \rangle\,\,.\label{Tij-gen}
\eea
 The hadron momentum $p=m_b v+k$ comprises  a residual component $k \simeq {\cal O}(\Lambda_{QCD})$. Redefining  $b(x)=e^{-i\,m_b v \cdot x} b_v(x)$ and writing $p_X=m_bv+k-q$, one has:
\be
(T^{ij})_{MN}=\langle H_b(v,s)|{\bar b}_v(0) \Gamma_M^{(i)\dagger} S_U(p_X) \Gamma_N^{(j)}b_v(0) |H_b(v,s)\rangle\,\, ,
\ee
with  $S_U(p_X)$  the $U$ quark propagator. The HQE  in powers of  $m_b^{-1}$  is carried out \cite{Chay:1990da,Bigi:1993fe} replacing $k \to iD$ ($D$ is  the QCD covariant derivative) and expanding
$S_U(p_X)=S_U^{(0)}-S_U^{(0)}(i {\spur D})S_U^{(0)}+S_U^{(0)}(i {\spur D})S_U^{(0)}(i {\spur D})S_U^{(0)}+ \dots\,\,$
where $S_U^{(0)}=\displaystyle\frac{1}{m_b {\spur v}-{\spur q} -m_U}$.
Writing $p_U=m_b v -q$, ${\cal P}=({\spur p}_U+m_U)$ and $\Delta_0=p_U^2-m_U^2$,  the expansion   reads:
\bea
\frac{1}{\pi}{\rm Im}(T^{ij})_{MN} &&= \frac{1}{\pi}{\rm Im}\frac{1}{\Delta_0}\langle H_b(v,s)|{\bar b}_v [\Gamma_M^{(i)\dagger} {\cal P} \Gamma_N^{(j)}]b_v |H_b(v,s) \rangle +\label{expansion}  \\
&&-\frac{1}{\pi}{\rm Im}\frac{1}{\Delta_0^2}\langle H_b(v,s)|{\bar b}_v[ \Gamma_M^{(i)\dagger} {\cal P}\gamma^{\mu_1}{\cal P} \Gamma_N^{(j)}](i D_{\mu_1})b_v |H_b(v,s)\rangle +\dots \nn
\eea
and  involves  the  matrix elements 
\be
{\cal M}_{\mu_1 \dots \mu_n}=\langle H_b(v,s)|({\bar b}_v)_a(i D_{\mu_1})\dots(i D_{\mu_n})(b_v)_b |H_b(v,s)\rangle ,\label{matel}
\ee
with $a,b$  Dirac indices.  These 
depend on  nonperturbative parameters: the matrix elements of the operators of increasing dimension in the HQET lagrangian. 
For example, ${\hat \mu}_\pi^2$ and ${\hat \mu}_G^2$ denote the matrix elements of the kinetic energy  and of the chromomagnetic operator, respectively.
The method proposed\cite{Dassinger:2006md} to compute  ${\cal M}_{\mu_1 \dots \mu_n}$  when $H_b$ is a meson has been generalized in the case of  baryons  for which the dependence on the spin $s_\mu$  in \eqref{matel} must be kept.\cite{Colangelo:2020vhu}
From the expressions of  ${\cal M}_{\mu_1 \dots \mu_n}$  the hadronic tensor  can be computed giving,  integrating  \eqref{diffwid},
double and single  decay distributions and  the full decay width  that can be  written as:
\be
\Gamma(H_b \to X \ell^- {\bar \nu}_\ell)=\frac{G_F^2 m_b^5 V_{Ub}^2}{192 \pi^3}\sum_i  \left\{C_0^{(i)}+\frac{\hat \mu^2_\pi}{m_b^2} C_{\mu_\pi^2}^{(i)}+\frac{\hat \mu^2_G}{m_b^2} C_{\mu_G^2}^{(i)}+\dots \right\} . \label{fullwidth}
\ee
The index  $i$ runs over the contribution of the various operators and of their interferences.
The   coefficients  $C^{(i)}$ depend on the NP couplings    in (\ref{hamil}).  Using \eqref{fullwidth} to fit experimental data allows the inclusive determination of $|V_{Ub}|$.
HFAG Collaboration\cite{HeavyFlavorAveragingGroup:2022wzx} provides as the average result using the inclusive procedure: $|V_{cb}|=(42.19 \pm 0.78)\times 10^{-3}$ with the   $b$ quark mass  in the  kinetic scheme.

\subsubsection{Possible solutions to the $V_{xb}$ puzzles}
Also for $V_{ub}$ inclusive and exclusive determinations are only marginally compatible: the HFAG averages  read
$|V_{ub}|_{\rm excl}=(3.67 \pm 0.09_{\rm exp} \pm 0.12_{\rm th}) \times 10^{-3}$  and 
$|V_{ub}|_{\rm incl}=(4.19 \pm 0.12 \pm^{0.11}_{-0.12}) \times 10^{-3}$.
The tension between the two values of  $|V_{ub}|$ and $|V_{cb}|$ is an old problem for which several explanations have been proposed.  
One is the violation of the quark-hadron duality assumption  underlying the inclusive determination.
 A consolidated idea was that NP could not be responsible of the tension. However, after the emergence of the anomalies in semileptonic $b \to c$ decays  reviewed below,   it appeared sensible  to relate these anomalies to the one  affecting $V_{cb}$.\cite{Colangelo:2016ymy}

Also  the correctness of procedure to extract $|V_{cb}|$ from  $B \to D^* \ell \bar \nu_\ell$ has been discussed.
In this case, the experimental determinations were based on the Caprini-Lellouch-Neubert (CLN) parametrization of the $B \to D^*$ form factors \cite{Caprini:1997mu}
 relying on the HQ symmetry relations, improved by  the inclusion of radiative and $1/m_Q$ corrections. However, 
using  the deconvoluted fully differential decay distribution for $B \to D^* \ell \bar \nu_\ell$ provided by Belle Collaboration\cite{Belle:2016kgw} and adopting the Boyd-Grinstein-Lebed
(BGL) parametrization of the FF\cite{Boyd:1995cf,Boyd:1995sq},   a value of $|V_{cb}|$  compatible
with the inclusive determination was found. \cite{Bigi:2017njr,Grinstein:2017nlq}
This solution was not confirmed by BaBar Collaboration that repeated the same kind of analysis adopting both sets of form factors and finding 
that the tension between the two determinations persists.\cite{BaBar:2019vpl}

A  recent study \cite{Martinelli:2023fwm}  reconsidered the role of  the FF. Adopting recent lattice results for these and new Belle data\cite{Belle:2023bwv} for semileptonic $B$ decays, they find  $R(D^*)$  compatible with data at 1.2 $\sigma$ level  and $|V_{cb}| $  in the range of the inclusive determination. 

\section{Flavour Anomalies}
The tension between inclusive/exclusive determinations of $|V_{ub}|$ and $|V_{cb}|$ and the Cabibbo anomaly are   examples of flavour anomalies, i.e. tensions between measurements and SM predictions. Other  deviations have been observed  with different levels  of significancy, even though  none of them can be recognized as an unambiguous  signal of NP. However, the number of observed deviations, together with  persistency of some of them might be hints to  BSM physics.
We consider below those 
 that have received more attention in recent years.

\subsection{FCNC decays induced by  $b \to s$ transition}
These processes are both loop and CKM suppressed in the SM, hence they are very sensitive to NP contributions. Several measurements of observables related to these transitions  have constrained a number of NP scenarios. However, a few anomalies have emerged that need to be scrutinized, in particular in the $b \to s \ell^+ \ell^-$ modes. In the SM these are described by the  effective Hamiltonian\cite{Buras:2020xsm}:
\be
H^{eff}=-\,4\,\frac{G_F }{\sqrt{2}} V_{tb} V_{ts}^*  \Big\{C_1 O_1+C_2 O_2 
+\sum_{i=3,..,6}C_i O_i+\sum_{i=7,..,10}C_i O_i   ] \Big\}\, . \label{hamil}
\ee
Terms proportional to $V_{ub} V_{us}^*$ are neglected.  $O_{1,2}$ are current-current operators while  $O_{i=3,\dots 6}$ correspond to QCD penguin operators:
\bea
O_1 &=& ({\bar c}_\alpha \gamma_\mu P_L \, b_\beta)({\bar s}_\beta \gamma^\mu P_L \, c_\alpha)\, , \hskip .3 cm  
O_2 = ({\bar c} \gamma_\mu P_L \, b)({\bar s} \gamma^\mu P_L \, c) , \label{O12}
\\
O_3&=&({\bar s} \gamma^\mu P_L \, b) \sum_q ({\bar q} \gamma^\mu P_L \, q) \, , \hskip .5 cm 
O_4=({\bar s}_\alpha \gamma^\mu P_L \, b_\beta) \sum_q ({\bar q}_\beta \gamma^\mu P_L \, q_\alpha)  ,\, \,\label{O34} \\
O_5&=&({\bar s} \gamma^\mu P_L \, b) \sum_q ({\bar q} \gamma^\mu P_R \, q) \, , \hskip .5 cm 
O_6=({\bar s}_\alpha \gamma^\mu P_L \, b_\beta) \sum_q ({\bar q}_\beta \gamma^\mu P_R \, q_\alpha) , \,\,\label{O56}
\eea
with  $P_{R,L}=\displaystyle\frac{1 \pm \gamma_5}{2}$,  $\alpha,\beta$  colour indices, $q=u,d,s,c,b$ in the sum in (\ref{O34})-(\ref{O56}).
$O_7, \,O_8$ are magnetic penguins; $O_9$, $O_{10}$ semileptonic electroweak penguin operators
\bea
O_7&=&\frac{e}{16 \pi^2} [{\bar s}\sigma^{\mu \nu}(m_s P_L + m_b P_R)\,b] F_{\mu \nu} \, , \label{O7} \\
O_8&=&\frac{g_s}{16 \pi^2}\Big[{\bar s}_{ \alpha} \sigma^{\mu \nu} \Big({\lambda^a \over 2}\Big)_{\alpha \beta} (m_s P_L + m_b P_R)  b_{ \beta}\Big] 
      G^a_{\mu \nu}  \, ,  \label{O8} 
\\
O_9&=&{e^2 \over 16 \pi^2}  ({\bar s} \gamma^\mu P_L \, b) \; {\bar \ell} \gamma_\mu \ell  \, , \hskip 1. cm
O_{10}={e^2 \over 16 \pi^2}  ({\bar s} \gamma^\mu P_L \, b) \; {\bar \ell} \gamma_\mu \gamma_5 \ell  \,.\label{O910}
\eea
$\lambda^a$ are the Gell-Mann matrices,  
$F_{\mu \nu}$ and $G^a_{\mu \nu}$  the
electromagnetic and  gluonic field strengths,  $e$ and $g_s$ the
electromagnetic and strong coupling constants, $m_{b,s}$  the $b$ and $s$ quark mass. 
Other operators can appear in BSM models  analogous to the previous ones but with opposite chirality or scalar, pseudoscalar or tensor operators.
In \eqref{hamil} a separation of scales is achieved:  long distance physics is encoded in the matrix elements of the operators, short distances in the coefficients that contain the information about the heavy degrees of freedom being integrated out. These could be SM particles or possible new particles: this is why  NP can
  modify the value of the  coefficients. 
 These processes are rare, with branching ratios predicted ${\cal O}(10^{-6})$ in the SM.  
Anomalies have been found in the measurements of several observables.\footnote{A recent review of experimental results can be found in \cite{Mathad:2023wpm}.}
The  branching fractions ${\cal B}(B \to K \mu^+ \mu^-)$ and ${\cal B}(B_s \to \phi \mu^+ \mu^-)$  display a deficit with respect to  the SM prediction. In $B \to K^* \mu^+ \mu^-$ one can define the quantity $P_5^\prime=J_5/(2\sqrt{-J_{2c}J_{2s}})$ where the $J_i$ are angular coefficients parametrizating  the fully differential decay rate for this mode. $P_5^\prime$ is a function of $q^2$, the lepton pair invariant mass. Data are sistematically higher than the SM prediction in the low $q^2$ bins. The significance of the deviation is at the level of $\simeq 3 \sigma$.\cite{LHCb:2020lmf}

In the case of the mode $B_s \to \mu^+ \mu^-$ the  measurement\cite{LHCb:2021vsc}  ${\cal B}(B_s \to \mu^+ \mu^-)=(3.9\pm^{0.46}_{0.43}\pm^{0.15}_{0-11})\times 10^{-9}$  agrees with the SM result\cite{Bobeth:2013uxa}  ${\cal B}(B_s \to \mu^+ \mu^-)=(3.66\pm 0.14)\times 10^{-9}$. We clarify below the importance of this decay.
Finally,   a recent Belle II result\cite{GlazovEPS}  indicates an excess, at $2.8 \sigma$  compared to  the SM, in the branching ratio of the decay $B^+ \to K^+ \nu \bar \nu$ that is also induced by  $b \to s $ quark level transition, but is governed by a different effective Hamiltonian.

The methodology of the  bottom-up approach can be explained in the case of these transitions. Starting from the effective Hamiltonian with all  possible NP operators a global fit is performed to find the values of the  coefficients that   reproduce the data. 
The various processes and observables have a different sensitivity to the  operators in the effective Hamiltonian. For example, $B_s \to \mu^+ \mu^-$ is important since it is sensitive only to the scalar and the axial vector operators and hence to NP models  introducing new scalar particles, such as those with more  Higgs doublets.

The methods to perform  the global fits    can differ in the number of  coefficients  included in the fit as well as in the statistical analysis. For example one can allow  only one coefficient  to deviate from the SM value  to understand the role of a single operator  before performing a fit with more parameters. Moreover, once a set of coefficients is found  suitable to reproduce data,   models   predicting coefficients in ranges different from those identified through the global fit could be discarded.

Many studies of $b \to s \ell^+ \ell^-$  modes in the top-down approach have been performed in view of their high sensitivity to NP. This has allowed to rule out some models and to constrain the parameter space of others. One example are models that add a new $U(1)$ component to the SM gauge group,  introducing a new neutral massive gauge boson $Z^\prime$. This may occur also as the result of the breaking of a larger group: an example will be given in section \ref{331}. An interesting case is when $Z^\prime$  can mediate FCNC at tree-level. Its contribution to a process like those considered in this section would be suppressed as $1/M_{Z^\prime}^2$ but could be competititve to the (loop-suppressed) SM one for suitable values of the  couplings of $Z^\prime$ to fermions.
Such couplings enter in the NP contribution to the Wilson coefficients and therefore are constrained by data: for example, data on $B_s \to \mu^+ \mu^-$ place bounds on $C_{10}^{NP}$. 

\subsection{
Charged current  processes induced by  $b \to c \ell \bar \nu_\ell$ transition}
In 2012 BaBar Collaboration measured
the ratios of  branching fractions ${\cal R}(D^{(*)})=\displaystyle\frac{{\cal B}(B \to D^{(*)} \tau \bar \nu_\tau)}{{\cal B}(B \to D^{(*)} \ell \bar \nu_\ell)}$ with $\ell=e,\,\mu$,
 in excess with respect to  the SM prediction, with a deviation  at $3.4\sigma$ level\cite{BaBar:2012obs}.
 Since then 
 Belle and LHCb Collaborations measurements, as well as new BaBar analyses,   confirmed the presence of the anomaly. The present average of  all  results\cite{HeavyFlavorAveragingGroup:2022wzx}: ${\cal R}(D)=0.298 \pm 0.004$ and ${\cal R}(D^*)=0.265 \pm 0.013$
  deviates at $3.2 \sigma$ from the SM prediction. 
  A  deviation is found\cite{LHCb:2017vlu} also in the ratio ${\cal R}(J/\psi)=\displaystyle\frac{{\cal B}(B_c^+ \to J/\psi \tau^+ \nu_\tau)}{{\cal B}(B_c^+ \to J/\psi \ell^+ \nu_\ell)}=0.71 \pm 0.17(\rm stat.)\pm0.18(\rm syst.)$ which is $2\sigma$ above the SM prediction. However, in this case the uncertainty related to the hadronic form factors is  much larger than for ${\cal R}(D^{(*)})$.
In the SM these are tree-level  processes mediated by the $W$ which couples with the same strenght to the three families of leptons. Consequently,
 these anomalies seem to indicate violation of the LFU in the third generation. LFU is one of the accidental symmetries of the SM.
 
Many  explanations for  these anomalies have been proposed, both in a bottom-up approach, both in  NP scenarios. 
Among the latter ones, the first considered candidate  was  the two Higgs doublet model.  In this framework a new scalar is introduced that, as  the SM Higgs, couples to fermions proportionally to their mass, hence providing a possible explanation for the enhancement of the semileptonic decay to $\tau$ leptons with respect to light leptons. However, in the simplest version of this model, only one new parameter is introduced, i.e. the ratio of the vacuum expectation values of the two Higgs doublets, and it could not be fixed in order to reproduce simultaneously the data for ${\cal R}(D)$ and ${\cal R}(D^*)$\cite{BaBar:2012obs}.

The observed  anomalies in $b \to c$ semileptonic exclusive decays of beauty mesons suggest   new analyses
of related processes involving other beauty hadrons, to enlarge the set of observables
useful  to test the SM predictions. 

The bottom-up approach offers a suitable framework for this purpose. 
The starting point is the generalized effective Hamiltonian Eq. \eqref{hamilgen}.  
One can check if for some values of the couplings it is possible to reproduce data and, in correspondence to such values, predict other observables induced by the same underlying transition. Similarly to  the case of the $b \to s$ modes,  one can allow only one coupling to vary or more than one simultaneously. 
We give a few examples of the two situations.

In the case of $b \to c$ semileptonic decays and, specifically, the mode ${\bar B}^0 \to D^{*+} \tau^- \bar \nu_\tau$, it can be shown that the inclusion of a new tensor operator in $H_{\rm eff}$ has an impact on the forward-backward asymmetry ${\cal A}_{FB}(q^2)$ that counts the difference between $\tau$ leptons emitted forward with respect to those emitted backward considering the $D^*$ direction of motion in the lepton pair rest-frame. This  quantity is a function of the lepton pair invariant mass $q^2$. The value  $q_0^2$ such that ${\cal A}_{FB}(q_0^2)=0$ is different in the SM and in the case where the new tensor operator is introduced.\cite{Biancofiore:2013ki,Colangelo:2018cnj}
\begin{figure}[b]
\begin{center}
\includegraphics[width=2.5in]{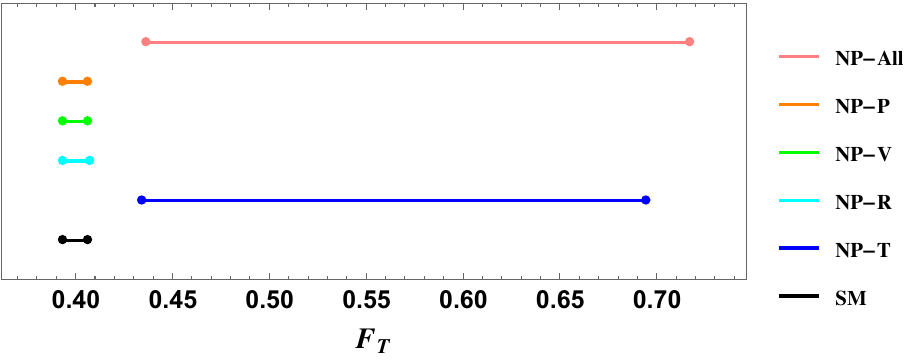}
\end{center}
\caption{Fraction of  transversely polarized  $B_d^*$. The lines correspond to   the SM,  to  the NP operators in Eq.~\eqref{hamil}  separately considered, and to the full set of NP operators. }
\label{figBcBdstar}
\end{figure}
The role of one operator can emerge also if other ones are taken into account.
This situation is examplified in Fig. \ref{figBcBdstar} that refers to the mode $B_c^+ \to B_d^* \ell^+ \nu_\ell$,  $\ell=e,\,\mu$. \cite{Colangelo:2021dnv} One can consider the fraction of $B_d^*$ that are produced with  longitudinal ($F_L$) or  transverse polarization ($F_T$). In the SM $F_T<0.5$, indicating that $B_d^*$ is produced dominantly with  a longitudinal polarization. The tensor operator can reverse the hierarchy. 
\begin{table}[t]
\tbl{Contribution of the operators in \eqref{hamilgen} to  various decays induced by $b \to u \ell \bar \nu_\ell$ transition.}
{\begin{tabular}{@{}ccccc@{}}
\toprule
 \hskip 3mm Mode  \hskip3mm & \hskip 3mm $\epsilon_V$ \hskip 3mm& \hskip 3mm  $\epsilon_S$  \hskip 3mm  &  \hskip 3mm   $ \epsilon_P$  \hskip 3mm &   \hskip 3mm    $\epsilon_T$   \hskip 3mm\\
\colrule
$B^- \to \ell^- \bar \nu_\ell$  & \checkmark & &  \checkmark& \\
 $B \to \pi \, \ell^- \bar \nu_\ell$  & \checkmark & \checkmark&  &  \checkmark  \\
 $B \to \rho \,  \ell^- \bar \nu_\ell$  &  \checkmark & & \checkmark  &  \checkmark \\
 $B \to a_1 \, \ell^- \bar \nu_\ell$  &\checkmark   & \checkmark &  & \checkmark \\
\botrule
\end{tabular}
}
\label{tbl1}
\end{table}
The role of the various operators can be inferred also considering that they contribute to a given process depending on the quantum numbers of the hadron in the final state. This is shown in Table \ref{tbl1} that  refers to  $b \to u \ell \nu_\ell$ transitions.\cite{Colangelo:2019axi} Noticeably, the couplings $\epsilon_S$ and $\epsilon_P$ are  complementary to each other.
\def\figsubcap#1{\par\noindent\centering\footnotesize(#1)}
\begin{figure}%
\begin{center}
  \parbox{2.1in}{\includegraphics[width=1.8in]{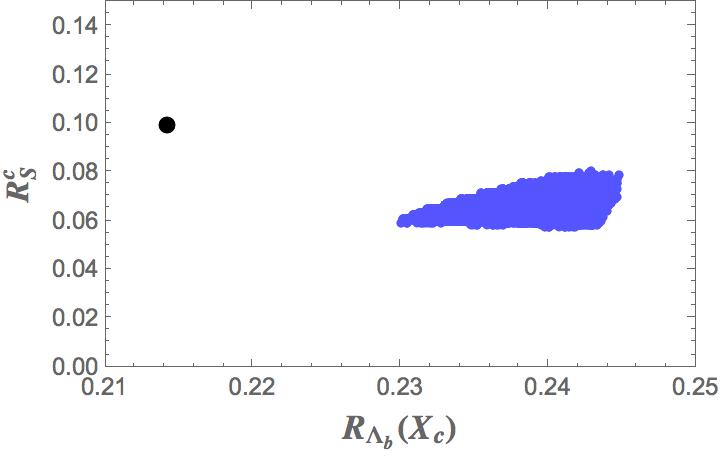}\figsubcap{a}}
  \hspace*{4pt}
  \parbox{2.1in}{\includegraphics[width=2.5in]{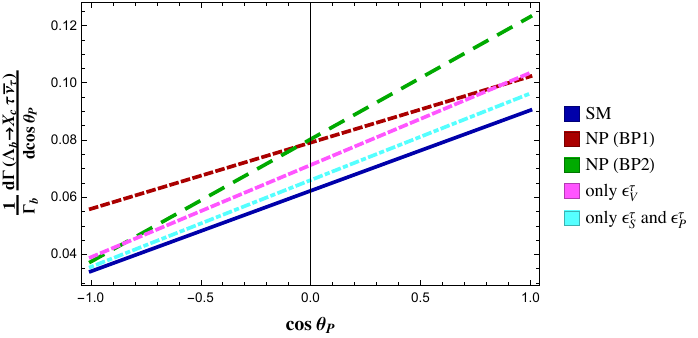}\figsubcap{b}}
  \caption{ (a) $\Lambda_b \to X_c \tau \bar \nu_\tau$:
 correlation between $R_{\Lambda_b}(X_c)$ and  $R_S^c$. The dot corresponds to SM, the blue region to NP. (b) $\dd \frac{1}{\Gamma_b} \frac{d \Gamma}{d\cos \theta_P}$  distribution  in SM and in  NP  with different   effective couplings.}%
  \label{figLambdab}
\end{center}
\end{figure}
In the bottom up approach a  possibility to find new observables that can probe LFU  is offered by semileptonic decays of beauty baryons such as $\Lambda_b$.  In the inclusive decay $\Lambda_b \to X_U \ell \bar \nu_\ell$, defining $\theta_P$ as the angle between the momentum $\vec p_\ell$ of the charged lepton and the direction $\vec s$ of the $\Lambda_b$ spin, one has that $\frac{d \Gamma (\Lambda_b \to X_U \ell \bar \nu_\ell)}{d {\rm cos} \theta_P}=A_\ell^U  +B_\ell^U \, {\rm cos} \theta_P$. The ratio of the intercepts for $\ell=\mu$ and $\ell=\tau$ ${\cal R}_{\Lambda_b}(X_U)=A_\tau^U/A_\mu^U$, analogous to ${\cal R}(D^{(*)})$, and the ratio of the slopes ${\cal R}_S^U=B_\tau^U/B_\mu^U$ represent two quantities that can probe LFU.\cite{Colangelo:2020vhu}  Fig.~\ref{figLambdab}(a) shows that   a  tensor operator sizably modifies these observables with respect to the SM, while  other operators have a reduced impact, as it can be inferred from the distribution in Fig.~\ref{figLambdab}(b).

\section{Working in  the top-down approach: 331 models}\label{331}
Many NP models  extend  the SM gauge group. Among these,   the 331 models\cite{Pisano:1991ee,Frampton:1992wt}
are based on  $SU(3)_c \otimes SU(3)_L \otimes U(1)_X$. This  is  first spontaneously broken to the SM   group $SU(3)_c \otimes SU(2)_L \otimes U(1)_Y$ and then   to $SU(3)_c \otimes U(1)_Q$. In such models 
the requirement of anomaly cancelation together with that of asymptotic freedom of QCD constrains the number of generations to be equal to the number of colours, hence answering the question of why three generations.
The same request imposes  that under  $SU(3)_L$ two quark generations  transform as triplets, one as an antitriplet. This is often chosen  to be the third generation, and the difference possibly at the origin of the large top mass.

The electric charge operator $Q$ is related to $T_3$ and $T_8$, two of the $SU(3)$ generators,  and to $X$, the generator of $U(1)_X$, through the relation
$Q=T_3+\beta T_8+X$. $\beta$ is a  parameter that defines a  variant of the model. 

The extension of the gauge group leads to introduce 5 new gauge bosons. Four of them are denoted by 
$Y^{Q_Y^\pm}$ and $V^{Q_V^\pm}$: their charges depend on the considered model variant. In all  variants,   a new neutral gauge boson $Z^\prime$ exists that mediates tree-level FCNC in the quark sector.  Its couplings to leptons are instead diagonal and universal.
The Higgs sector  consists of three $SU(3)_L$ triplets and one sextet.
In the fermion triplets, new heavy fermions appear together with  the SM ones.

As in  SM, two unitary rotation matrices $U_L$ (for up-type quarks) and $V_L$ (for down-type ones) transform flavour eigenstates into quark mass eigenstates and  one has   $V_{CKM}=U_L^\dagger V_L$. However,  in  SM $V_{CKM}$ appears only in charged current interactions and $U_L,\,V_L$ do not  appear individually. Instead, in 331 models only one matrix  between $U_L$ and $V_L$ can be expressed in terms of $V_{CKM}$ and the other one. The remaining rotation matrix is usually chosen to be $V_L$. It  enters in the $Z^\prime$ couplings to quarks, and it  is parametrized 
in terms of three angles $\tilde \theta_{ij}$ ($i,j=1,2,3$) with sines and cosines denoted ${\tilde c}_{ij},\,{\tilde s}_{ij}$, and three phases $\delta_i$ ($i=1,2,3$).
The Feynmann rules for $Z^\prime$ couplings to quarks\cite{Buras:2012dp}, show 
that the $b \to d$ transition, relevant for the $B_d$ system, involves only the parameters ${\tilde s}_{13}$ and $\delta_1$, while the $B_s$ system depends  on 
${\tilde s}_{23}$ and $\delta_2$.\cite{Buras:2012dp} A peculiar feature  is that also the $s \to d$ transition, relevant for the kaon sector, and the $c \to u$ one, relevant for charm  processes, depend on ${\tilde s}_{13}$, ${\tilde s}_{23}$ and $\delta_2$, $ \delta_1$. \cite{Colangelo:2021myn,Buras:2021rdg}
Hence, 
stringent correlations among flavour observables  are predicted, they can discriminate this model from others.
In order to derive such correlations in a given 331 variant one preliminarly finds allowed  regions, {\it oases}, for the four parameters ${\tilde s}_{13}$, ${\tilde s}_{23}$, $\delta_2$, $ \delta_1$ in which experimental data on selected flavour observables are reproduced within their uncertainty. These are usually chosen to be  the mass differences $B_d-{\bar B}_d$, $B_s-{\bar B}_s$, $K^0-{\bar K}^0$;   the CP asymmetries $S_{J/\psi \,K_S}$ in the mode $B^0 \to J/\psi K_S$ and $S_{J/\psi \, \phi}$ in $B_s \to J/\psi \phi$; the parameter $\epsilon_K$ related to CP violation in the kaon system.  In the SM  these observables are related to the neutral meson oscillations  described through box diagrams. 
In 331 models a new contribution is given by the tree-level diagram mediated by
 $Z^\prime$. 
 
\begin{figure}
\begin{center}
\includegraphics[width=2.8in]{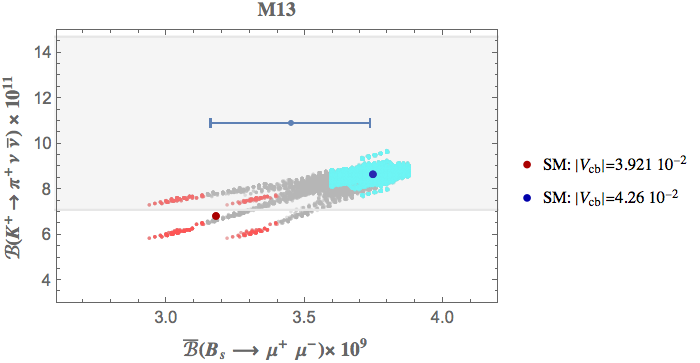}
\caption{Correlation between $ {\cal B}(K^+ \to \pi^+ \nu  \bar \nu) $ and ${\bar {\cal B}}(B_s \to \mu^+ \mu^-) $. The gray points span all the allowed parameter space.
    The red ones correspond to $|V_{cb}| \in [0.0386,\,0.0398]$, the cyan  to $|V_{cb}| \in [0.0422,\,0.043]$. The SM results are shown for two values of $|V_{cb}|$   in the legenda. The light gray region and the blue range are
the experimental ranges of ${\cal B}(K^+ \to \pi^+ \nu  \bar \nu) $ and ${\bar {\cal B}}(B_s \to \mu^+ \mu^-) $.}
\label{331correlation}
\end{center}
\end{figure}

The SM prediction for 
 a given flavour observable depends on the CKM parameters. Choosing  $|V_{cb}|$ and $|V_{ub}|$ as two of the independent parameters of $V_{CKM}$ one can  study correlations among flavour observables in selected ranges of these   CKM elements. 
Fig. \ref{331correlation} shows the  correlation  between 
$ {\cal B}(K^+ \to \pi^+ \nu  \bar \nu) $ and ${\bar {\cal B}}(B_s \to \mu^+ \mu^-) $ in one 331 variant.\cite{Buras:2023ldz} The inclusive values of $|V_{cb}|$ correspond to points that can be compatible with the experimental result for ${\bar {\cal B}}(B_s \to \mu^+ \mu^-) $ performing slightly better than the SM  corresponding to $ {\cal B}(K^+ \to \pi^+ \nu  \bar \nu)\le 10^{10} $.
Exclusive values of $|V_{cb}|$  produce values of ${\bar {\cal B}}(B_s \to \mu^+ \mu^-) $ and $ {\cal B}(K^+ \to \pi^+ \nu  \bar \nu) $ simultaneously smaller than the experimental range, a prediction that can be used to test this variant.

\section{Conclusions and perspectives}
Many novelties  enrich the  panorama of flavour physics. Among these, the  flavour anomalies  prompting  efforts to reveal  NP.  Besides those discussed in this lecture, other important ones are the anomalous magnetic moment of the muon and the related problem of the hadronic $e^+ e^-$ cross section; the  CP violating parameter $\epsilon^\prime/\epsilon$  in kaon decays. Other investigations should be pursued  looking for processes forbidden in the SM as  lepton decays  $\mu^- \to e^- \gamma$, $\mu^- \to e^- e^+ e^-$.
Remarkably  flavour  observables can access NP scales  higher than direct searches foreseen in the near future \cite{EuropeanStrategyforParticlePhysicsPreparatoryGroup:2019qin}
 confirming this sector as very promising to discover BSM physics. 

\section*{Acknowledgments}
I am grateful to the Directors of the school: Prof. A. Zichichi and Prof. A. Zoccoli and to the Directors of the course: Prof. A. Bettini and Prof. A. Masiero for inviting me to give this lecture and to the EMFCSC Director: Dr. F. Ruggiu for the perfect organization of the school. I thank A.J. Buras, P. Colangelo,  F. Loparco and N. Losacco for collaboration on 
some of the topics  covered in my lecture.\\
My research is supported by the INFN {\it Iniziative Specifiche} QFT-HEP and SPIF.

\bibliographystyle{ws-procs961x669}
\bibliography{defazio}

\end{document}